\documentclass[10pt,pre,aps,twocolumn,superscriptaddress,floatfix,notitlepage]{revtex4-2}
\usepackage[latin1]{inputenc}
\usepackage{hyperref}
\hypersetup{
    colorlinks=true,
    linkcolor=blue,
    filecolor=magenta,      
    urlcolor=cyan,
    pdftitle={Autonomous quantum clocks using athermal resources},
    pdfpagemode=FullScreen,
    }
\usepackage{mathrsfs,dsfont}
\usepackage{amsmath}
\usepackage{amsfonts,natbib}
\usepackage{amssymb}
\usepackage{bm}
\usepackage[pdftex]{graphicx}

\usepackage{graphicx}

\begin{document}
	\title{Autonomous quantum clocks using athermal resources} 
	\author{Sreenath K. Manikandan}
	\email{sreenath.k.manikandan@su.se}
	\affiliation{Nordita,
KTH Royal Institute of Technology and Stockholm University,
Hannes Alfv\'{e}ns v\"{a}g 12, SE-106 91 Stockholm, Sweden}
\date{\today}
	\begin{abstract}
Here we explore the possibility of precise time-keeping in quantum systems using athermal resources. We show that quantum measurement engineered reservoirs can be used as athermal resources to drive the ticks of a quantum clock.   Two and three level   quantum systems act as transducers in our model, converting the quantum measurement induced noise to produce a series of ticks. The ticking rate of the clock is maximized when the measured observable maximally non-commutes with the clock's Hamiltonian.   We use the large deviation principle to characterize the statistics of observed ticks within a given time-period and show that it can be sub-Poissonian---quantified by Mandel's Q parameter---alluding to the quantum nature of the clock.    
We discuss the accuracy and efficiency of the clock, and extend our framework to include hybrid quantum clocks fueled by both measurements, and thermal resources. We make comparisons to relatable recent proposals for quantum clocks, and discuss alternate device implementations harvesting the quantum measurement engineered non-equilibrium conditions,  beyond the clock realization.  
	\end{abstract}
	\maketitle
\section{Introduction}The thermodynamic cost of precise time-keeping has become an interesting direction of research in the quest for autonomous quantum clocks~\cite{erker_autonomous_2017,milburn_thermodynamics_2020,pearson_measuring_2021,woods_autonomous_2019,woods_autonomous_2021,schwarzhans_autonomous_2021}. It is a problem relevant to fundamental physics as well, since ideal clocks are also systems that break continuous time-translational symmetry~\cite{shapere_classical_2012,wilczek_quantum_2012}. Probing the limits of achieving precise time-keeping in quantum systems therefore paves way to better understand the notion of time-translational symmetry breaking, and the emergence of possible time-crystalline phases of matter~\cite{shapere_regularizations_2019,sacha_time_2017}. This is also a field of much debate, as there are no---go theorems that exclude the existence of ideal quantum clocks~\cite{pauli_general_1980,pauli_allgemeinen_1933}. Notable earlier works on quantum mechanics of clocks include Hamiltonian description for quantum clocks, discussed in~\cite{peres_measurement_1980}. More recently, axiomatic approaches for studying quantum clocks have been presented in refs.~\cite{woods_quantum_2022,woods_autonomous_2019,woods_autonomous_2021}. Precise measurements of time or space-time distances and synchronization of quantum clocks are also exciting avenues of research in gravity~\cite{salecker_quantum_1997,jozsa_quantum_2000,smith_quantum_2020}. 
\begin{figure}[b]
    \includegraphics[width=0.9\linewidth]{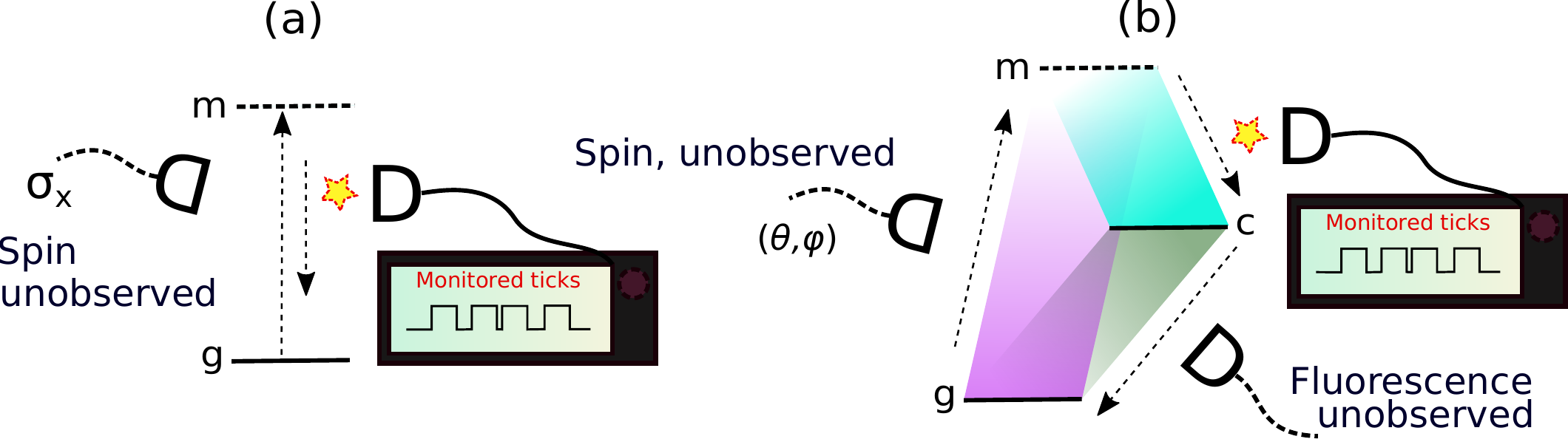}
    \caption{Ignorant observers can make a quantum clock tick. We envision (a) two level and (b) three level quantum systems as clock devices, using unobserved quantum measurements of spin as a fuel. The clock ticks are registered by measuring fluorescence emission. We also consider hybrid quantum clock realizations combining measurement and thermal resources.} 
    \label{fig:main}
\end{figure}

Here we approach the problem from engineering a clock device perspective: a device which can provide a time reference, or estimate the time elapsed for processes that occur at a rate slower than the ticking rate of the clock, $\gamma_{\text{tick}}$. Such an approach was recently proposed for thermal quantum clocks in~\cite{erker_autonomous_2017}, and experimentally investigated in~\cite{pearson_measuring_2021}. The idea therein is that a clock is a thermodynamic device, at the heart of which is a heat engine which delivers the work that keeps the clock ticking. The performance of the engine is limited by thermodynamic principles, which manifestly prevents the existence of a perpetual ticking clock. When the ticks are generated by coupling a quantum system to both hot and cold reservoirs, it can be viewed as a thermal  quantum clock; one may therefore provide a thermodynamic characterization of the quantum clock by extending the thermodynamic principles of quantum engines. The clock in ref.~\cite{erker_autonomous_2017} for example, produces ticks by rectifying thermal fluctuations alone, and generates some residual heat dumped into an ambient cold reservoir. Crucially, the clock does not require another clock in its dynamical description, making it an autonomous thermal quantum machine (also see~\cite{mitchison_quantum_2019}). 

Thermal clocks also inherit the limitations of other thermal machines, for example, they do not produce ticks at zero temperature. More generic examples of clocks realizing a limit cycle with alternate fueling mechanisms are discussed by Milburn in~\cite{milburn_thermodynamics_2020}, which also motivates to think about comparable simple models for clocks in the quantum regime, where the resources utilized are autonomous, but athermal. By athermal, we refer to scenarios where the non-equilibrium conditions relevant for steady state machines are not realized by an explicit temperature gradient. Principles of statistical physics and quantum mechanics allow for fundamental sources of randomness other than thermal resources; a canonical example of an athermal resource offered by quantum mechanics, well studied recently in this spirit, is the quantum measurement process~\cite{dressel_arrow_2017,manikandan_fluctuation_2019,harrington_characterizing_2019,jayaseelan_quantum_2021,elouard_role_2017,levy_quantum_2016}. Individual realizations of time-continuous quantum measurements~\cite{wiseman_quantum_2010,chantasri_stochastic_2015,weber_mapping_2014,chantasri_action_2013,katz_reversal_2008} induces stochastic quantum dynamics in the Hilbert space of the quantum system, exchanging energy and entropy with the system of interest. The average dynamics of the quantum system when the measurement outcomes are not read corresponds to the dynamics of an open quantum system with the crucial difference that the reservoir in this case is an engineered one: the quantum measurement apparatus. For examples of these devices, please see refs.~\cite{levy_quantum_2016,elouard_efficient_2018,elouard_extracting_2017,manikandan_efficiently_2022,buffoni_quantum_2019,elouard_role_2017,yi_single-temperature_2017,ding_measurement-driven_2018,son_monitoring_2021,bhandari_continuous_2021}. The interesting generalization we explore in the present article is that different quantum measurement interactions---such as for the measurements of quantum spin and fluorescence---maybe combined to realize non-equilibrium steady states in a three-level quantum system.  The non-equilibrium situation maybe taken advantage of for a class of quantum devices, including simple models for quantum clocks that we highlight in the article. The model is also extended further to include thermal resources, towards realizing a hybrid quantum clock. A different class of measurement driven clocks with a quantum two level system were also discussed recently in~\cite{he_measurement_2022}, which we make brief comparisons to, later in the discussions.  

The simplest descriptions possible for measurement apparatuses in condensed matter platforms reduces to a steady stream of electron current in the neighborhood of the quantum system, owing to a chemical potential gradient, such as a quantum point contact that responds to charge~\cite{korotkov_continuous_1999}. Therefore the athermal quantum clock that we propose in this article can be thought of as a minimal model for clocks driven by a chemical potential gradient, equivalent to a battery-driven clock. Fundamentally, the chemical potential gradients which drive the measurement apparatuses at zero temperature makes the system non-equilibrium, similar to thermal clocks where the non-equilibrium conditions are realized by a temperature gradient. At the quantum level, they are individual electrons traversing the point contact at random frequencies that depend on the properties of the quantum point contact, but effectively the macroscopic observables such as the electron current and their fluctuations determine the measurement characteristics of the apparatus. When coupled to the system of interest without additional readout capabilities, the measurement apparatus acts like an additional reservoir in the problem in the canonical sense, where the timescale of interaction with the apparatus characterized by a measurement rate, $\gamma_{m}$.   In the above sense, we call a measurement apparatus an autonomous resource; when the measurement outcomes are not read, the apparatus enter the dynamics in the same footing as thermal resources, as additional terms in the Lindbladian with a rate $\gamma_{m}$.  

For athermal quantum clocks, we consider   two and three level quantum systems\footnote{Three level quantum systems have been used as prototypical examples for quantum thermal machines since the earlier days of quantum thermodynamics, see for example, refs.~\cite{scovil_three-level_1959,geusic1959three}.}   coupled to athermal reservoirs---engineered via quantum measurements of spin and fluorescence---as the clock device. See Fig.~\ref{fig:main}.  The two and three level quantum systems   essentially act like transducers which convert the measurement noise into recordable ticks of the clock.      
 The quantum nature of the clocks manifests in two ways: (1) measurements in a basis that non-commutes with the clock's Hamiltonian is necessary for generating the clock ticks, and (2) the statistics of the number of ticks generated in a given time-period can be sub-Poissonian in resonance fluorescence~\cite{mandel_sub-poissonian_1979}, suggesting the feasibility of more regular (than Poissonian) clocks in the quantum regime. An added benefit is that the finite-time statistics of ticks is often exactly solvable using the large deviation principle (see for example, Ref.~\cite{garrahan_thermodynamics_2010}), and we present the two level example to primarily demonstrate this. This simple model also captures the essential details of the analysis that follows for a three level quantum system, which also incorporates additional dissipation.    
 
A possible interest beyond the condensed matter paradigm is that quantum measurement maybe implemented in a variety of different ways, and invites additional questions of fundamental interest such as what would be the back-action of observing ticks of the clock, on the quantum dynamics of the clock itself. Given the relevance of this question also for thermal quantum clocks, we envision hybrid quantum clocks, which are fueled by both measurements, and thermal resources. The framework discussed here also opens ways to address a variety of relatable questions which are accessible in experiments, thereby opening additional avenues for exploring the nature of clocks in quantum mechanics, in addition to the practical interests in precise time-keeping.

This article is organized as follows. 
We describe the dynamics of two level system subject to engineered measurement interactions in Sec.~\ref{twolevel}. We also derive the exact statistics of ticks observed within a given finite but large time-period using the principle of large deviation~\cite{garrahan_thermodynamics_2010}. We subsequently characterize the three level quantum clock in Sec.~\ref{threelevel}.   A natural extension to hybrid quantum clocks by combining measurements and thermal resources is discussed in Sec.~\ref{hybrid}. We conclude by discussing the implications of this work and some future directions, in Sec.~\ref{discuss}.

\section{Dynamics of athermal quantum clocks: A two level example\label{twolevel}}

\begin{figure}
    \centering
\includegraphics[width=\linewidth]{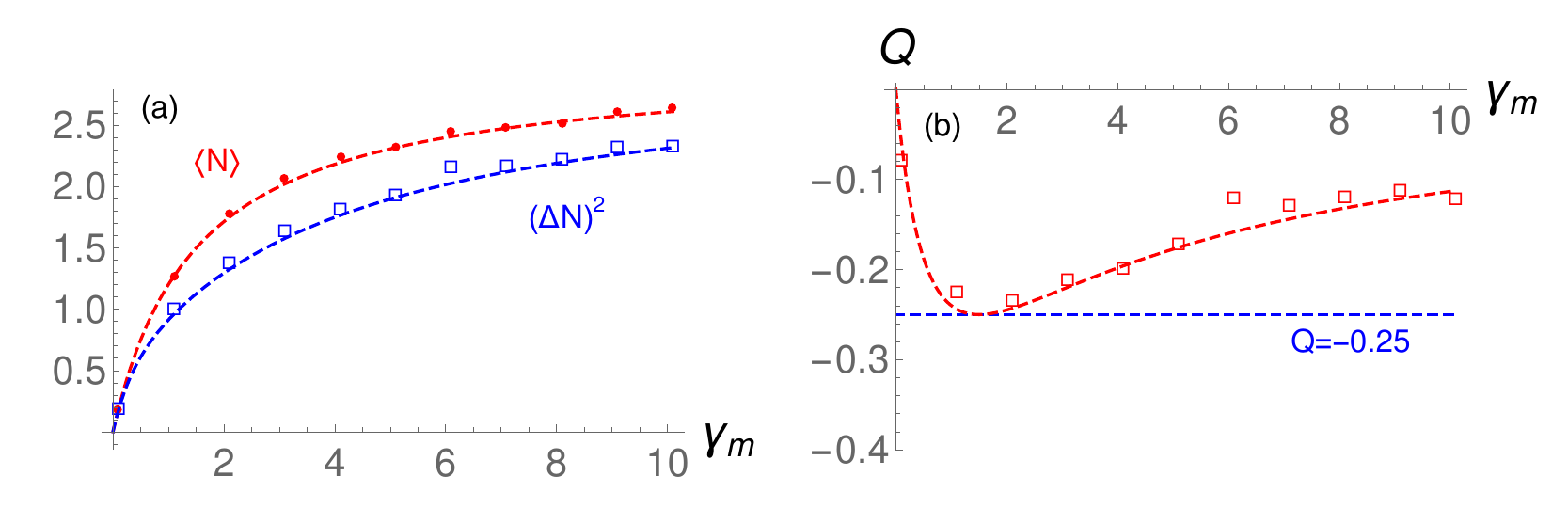}
    \caption{Characterizing the athermal quantum clock in a two level system: (a) Mean and variance of clock ticks. The dashed lines are predictions from using the large deviation principle, discussed in Sec.~\ref{twolevel}. We use $\Omega=1, \gamma_{w}=6$. In simulations, we generate $5\times 10^3$ number of trajectories using a number of time steps $n=1000$ with $dt=10^{-3}$. The system is initialized in the steady state.  (b) The Mandel Q parameter, with the achievable minimum equal to -0.25. The dashed lines are predictions based on large deviation principle.}
    \label{fig:athermaltwolevel}
\end{figure}

    We begin by considering a two level quantum system subject to spin and fluorescence measurement as clock device. The Hamiltonian is given by $\hat{H}=\Omega|e\rangle\langle e|,$ where $|e\rangle$ is the excited state of the two level quantum system. Further, we consider time continuous, and weak spin measurements of the observable $\sigma_{x}=|e\rangle\langle g|+|g\rangle\langle e|$ of the qubit, at a rate $\gamma_{m}$. We use the operator-sum representation (also known as the Kraus representation)~\cite{sudarshan_stochastic_1961,jordan_dynamical_1961,kraus_general_1971,kraus1983states} to describe and simulate the measurement dynamics. The spin measurement operators can be mapped to the following form   (see App.~\ref{KrausSpin} for details)  ,
\begin{equation}
    \hat{M}(r)=(4\epsilon_{m})^{1/4}\exp[-2\epsilon_{m}(r\hat{\mathbb{I}}-\sigma_{x})^{2}]/\pi^{1/4},
\end{equation}
where $\epsilon_{m}=\gamma_{m}dt$. The spin measurements, when unobserved, serves as the pumping process which continuously adds energy into the system. By closely following the discussion in~\cite{jacobs_straightforward_2006}, the measurement readout $r$ can be approximated by a stochastic signal $r\approx \langle \sigma_{x}\rangle +dW/(\sqrt{8\gamma_{m}}dt)$, where $dW$ is a Gaussian random variable (Wiener noise) having variance $dt$ and mean zero.  The measurement operators satisfy the completeness relation, $\int \hat{M}(r)^{2} dr =\hat{\mathds{1}}$. The quantum state evolution subject to unobserved spin measurements can be approximated by Taylor expanding the density matrix to first order in $\epsilon_{m}$, which requires one to keep terms of order $dW\sim\sqrt{dt}$ and $dW^{2}\sim dt$, and then setting terms that are proportional to $dW$ equal to zero. One obtains the simple rule, $d\rho = -\epsilon_{m}[\sigma_{x},[\sigma_{x},\rho]]$~\cite{jacobs_straightforward_2006}.  The effect of unobserved spin measurements can also be thought of as an additional Hermitian dissipator in the problem, $L=\sqrt{2}\sigma_{x}$. This also hints at the generic nature of the construction, as the measurement of a given observable can be implemented in a variety of different ways that are mathematically equivalent.  

The qubit fluorescence is also observed simultaneously via photon counting, and this serves as the clock signal. The Kraus operators describing the process are given by (see Refs.~\cite{lewalle_measuring_2020,jordan_anatomy_2016} for a review),
\begin{equation}
    \hat{M}_{w}(0)=\left(
\begin{array}{ccc}
  \sqrt{1-\epsilon_{w}} & 0 \\
 0 & 1 \\
\end{array}
\right),~\hat{M}_{w}(1)=\left(
\begin{array}{ccc}
  0 & 0 \\
 \sqrt{\epsilon_{w}} & 0 \\
\end{array}
\right),
\end{equation}
where $\epsilon_{w}=\gamma_{w}dt$, $\gamma_{w}$ the measurement rate, and $0$ and $1$ correspond to the number of photons detected between time $t$ and $t+dt$. The ordering of basis vectors assumed is $(|e\rangle, |g\rangle)$, such that these Kraus operators represent an amplitude damping channel.

The dynamics of the three level system subject to observation of ticks (or no ticks) can be modeled as follows. At any value of coordinate time $t$, a tick is recorded with probability,
\begin{eqnarray}
    p_{\text{tick}}&=&\text{tr}\bigg[\hat{M}_{w}(1)\bigg(\int dr\hat{\mathcal{M}}_{dt}(r)\rho(t)\hat{\mathcal{M}}_{dt}(r)^{\dagger}\bigg)\hat{M}_{w}(1)^{\dagger}\bigg],\nonumber\\
\end{eqnarray}
and no tick is observed with probability, $1-p_{\text{tick}}$. We have denoted $\hat{\mathcal{M}}_{dt}(r)=\hat{M}(r)e^{-i\hat{H}dt}$. Since the observation of ticks is probabilistic ($k=0$ for no ticks, $k=1$ for observing a tick), we have the following conditional evolution of the system at every time $t$,
\begin{equation}
    \rho(t+dt) = \frac{\hat{M}_{w}(k)\big[\int dr\hat{\mathcal{M}}_{dt}\rho(t)\hat{\mathcal{M}}_{dt}^{\dagger}\big]\hat{M}_{w}(k)^{\dagger}}{\text{tr}\big\{\hat{M}_{w}(k)\big[\int dr\hat{\mathcal{M}}_{dt}\rho(t)\hat{\mathcal{M}}_{dt}^{\dagger}\big]\hat{M}_{w}(k)^{\dagger}\big\}}.
\end{equation}
    In simulations, we approximate $\int dr\hat{\mathcal{M}}_{dt}\rho(t)\hat{\mathcal{M}}_{dt}^{\dagger}\approx\rho(t)-dt \big(i[\hat{H},\rho(t)]+\gamma_{m}[\sigma_{x},[\sigma_{x},\rho(t)]]\big)$. 

The statistics of the total number of ticks observed in a finite duration
 can be computed from the finite-time moment generating function of the number of counts within a given period. We use the large deviation principle to approximate the moment generating function in the large but finite-time limit, which can be identified as the largest eigenvalue of the the tilted Lindblad generator for the process~\cite{garrahan_thermodynamics_2010,touchette_large_2009}. The tilted Lindblad generator for the present example has the following form,
\begin{eqnarray}
    \mathcal{W}(s)[\rho] &=& -i[H,\rho]-\gamma_{m}[\sigma_{x},[\sigma_{x},\rho]]+\gamma_{w}[e^{-s}\sigma_{-}\rho\sigma_{+}\nonumber\\&-&(\sigma_{+}\sigma_{-}\rho+\rho\sigma_{+}\sigma_{-})/2].
\end{eqnarray}
When  the tilt parameter    $s=0$, we obtain the steady state of the system as solution to $\frac{d\rho}{dt} = \mathcal{W}(0)[\rho]$. The steady state for the present example has no coherences, and is characterized by $P_{e} = \rho_{ee}^{s} = 2\gamma_{m}/(4\gamma_{m}+\gamma_{w})$. 

The largest real eigenvalue of the tilted Lindblad generator is given by,
\begin{eqnarray}
  &&\Theta(s)=e^{-s}/2\times \nonumber\\&&   \left[\sqrt{e^s \left(e^s \left(16 \gamma_{m}^2+\gamma_{w}^2\right)+8 \gamma_{m} \gamma_{w}\right)}-e^s (4 \gamma_{m}+\gamma_{w})\right].\nonumber\\
\end{eqnarray}
The average number of ticks observed in any finite duration $t$ is given by,
\begin{equation}
    \langle N\rangle  = -\Theta'(s)|_{s=0}t = P_{e}\gamma_{w}t=\gamma_{\text{tick}}t.
\end{equation}
We compare to simulations in Fig.~\ref{fig:athermaltwolevel}(a). We note that the average number of ticks (or the observable clockwork) can be predicted from knowing the steady state of the dynamics and fluorescence decay rate. Computing the variance of ticks, on the other hand, require additional considerations. The variance of ticks is given by,
\begin{eqnarray}
    &&\langle N^{2}\rangle-\langle N\rangle^{2}  =\Theta''(s)|_{s=0}t\nonumber\\
    &&=\frac{2 \gamma_{m} \gamma_{w} \left(16 \gamma_{m}^2+4 \gamma_{m} \gamma_{w}+\gamma_{w}^2\right)}{\left[(4 \gamma_{m}+\gamma_{w})^2\right]^{3/2}}t.
\end{eqnarray}
The sub-Poissonian nature of ticks can be read out from the Mandel's Q parameter~\cite{mandel_sub-poissonian_1979}. For our example, the Q parameter is found to be,
\begin{equation}
   Q=-\Theta''(s)|_{s=0}/\Theta'(s)|_{s=0}-1 = -\frac{4 \gamma_{m} \gamma_{w}}{(4 \gamma_{m}+\gamma_{w})^2}<0.
\end{equation}
The negative value of Q indicates sub-Poissonian statistics, and the quantum nature of observed ticks. The best Q that can be realized in the problem is $Q=-1/4$, which happens when $\gamma_{w}=4\gamma_{m}$. See Fig.~\ref{fig:athermaltwolevel}(b), where we compare the predictions based on large deviation theory to numerical simulations. The clock can therefore ticks more regualarly than a Poissonian clock (see Ref.~\cite{milburn_thermodynamics_2020}), and fundamentally, this occurs because once a tick has happened (i.e., a quantum jump has occured), the probability of another jump instantaneously again is zero. The probability of a subsequent jump increases with passing time though, since the time-continuous quantum measurements of $\sigma_{x}$ populates the excited state partially via continuous pumping.

We may write the number of ticks obtained in any working duration $t_w$ as,
    \begin{eqnarray}
        N(t_w)&=&\langle N(t_w)\rangle\pm\delta N(t_w)\nonumber\\&=&\gamma_{\text{tick}}t_w\pm \sqrt{(1+Q)\gamma_{\text{tick}}t_w}.
    \end{eqnarray}
  
Therefore an interesting quantity for the quantum clock implementation is the relative error in estimating the number of ticks in a given working duration $t_w$, given by,
\begin{equation}
    \frac{\delta N(t_w)}{\langle N(t_w)\rangle}=\sqrt{(1+Q)}/\sqrt{\gamma_{\text{tick}}t_{w}},
\end{equation}
which indicates that the relative error in measuring time decreases with increasing the duration one intends to measure. Since $-1/4\leq Q<0$ for the clock model we described above, the relative error for the quantum clock described here is smaller than a Poissonian clock as well for which $Q=0.$ 

The two level example discussed here serves to highlight the essential aspects of producing clock ticks, as well as characterising its performance. It also had the advantage that the statistics is exactly solvable in the large but finite time limit. We now proceed to incorporate some dissipation into the clock model, by considering a three level quantum system.

\section{A three level quantum system as the clock device\label{threelevel}}

The three levels are labelled by $|m\rangle,|c\rangle$ and $|g\rangle$. The Hamiltonian is given by,
\begin{equation}
    \hat{H}=\Omega_{m} |m\rangle\langle m|+\Omega_{c} |c\rangle\langle c|.
\end{equation}
Note that the free evolution does not create any coherences in the quantum system. We assume $\Omega_{m}>\Omega_{c}$ and that the frequency $\Omega_{w}=\Omega_{m}-\Omega_{c}$ is accessible experimentally. On the other hand, we assume the frequency $\Omega_{c}$ corresponds to inaccessible losses in the system to the ambient cold reservoir via spontaneous emission at a rate $\gamma_{c}$. The ambient temperature is also assumed very low ($k_{\text{B}}T_{c}\ll \Omega_{c}$) such that it can be modeled as a zero temperature reservoir. We generalize the description to arbitrary, but finite temperature of the reservoirs in Sec.~\ref{hybrid}. 

\subsection{The continuous pumping of excitations\label{pumping}}

We consider generalized weak quantum spin measurements in the basis, $\{|\theta,\phi,+\rangle=\cos{(\theta/2)}|m\rangle +\sin{(\theta/2)}e^{-i\phi}|g\rangle, |c\rangle, |\theta,\phi,-\rangle=\sin{(\theta/2)}e^{i\phi}|m\rangle -\cos{(\theta/2)}|g\rangle\}$ that pump excitation between levels $g$ and $m$. In the Hilbert space of the three level system, this corresponds to measuring the observable,
\begin{equation}
\hat{X}=|\theta,\phi,+\rangle\langle \theta,\phi,+|-|\theta,\phi,-\rangle\langle \theta,\phi,-|.
\end{equation}
The observable has eigenvalue zero for the state $|c\rangle$, which means that the state is essentially a dark state in the quantum optics sense as far as the spin measurement is concerned. We may also evaluate the commutator, $[\hat{X},\hat{H}] = i[i\Omega_{m} e^{i\phi}\sin(\theta)|m\rangle\langle g|+\text{H.c.}]$, which indicates that the clock Hamiltonian maximally non-commutes with the measured quantum spin observable when $\theta=\pi/2$. 

 To describe the measurement process in a time-continuous manner, we again use Gaussian measurement operators describing the measurement process. Physically, one can think of a stream of photons in a Gaussian wavepacket, undergoing momentary collisional interactions with the three level quantum system via a dispersive ($\propto\hat{a}^{\dagger}\hat{a}\otimes\hat{X}$) coupling, where the photons do not interact with excitations in the $|c\rangle$ level. The measurement operators can be mapped to the following form,
\begin{equation}
    \hat{M}(r)=(4\epsilon_{m})^{1/4}\exp[-2\epsilon_{m}(r\hat{\mathbb{I}}-\hat{X})^{2}]/\pi^{1/4},
\end{equation}
where $\epsilon_{m}=\gamma_{m}dt$. The measurement readout $r$ is represented by a stochastic signal $r\approx \langle \hat{X}\rangle +dW/(\sqrt{8\gamma_{m}}dt)$.  The measurement operators again satisfy the completeness relation, $\int \hat{M}(r)^{2} dr =\hat{\mathds{1}}$. The change in the density matrix subject to unobserved spin measurements can be approximated by, $d\rho = -\epsilon_{m}[\hat{X},[\hat{X},\rho]]$~\cite{jacobs_straightforward_2006}, which is equivalent to having an additional Hermitian dissipator in the problem, $L=\sqrt{2}\hat{X}$. 
\subsection{Inaccessible losses}
   The inaccessible losses in the system to a zero-temperature bath can be modeled as an unobserved fluorescence measurement interaction, whose rate may be determined by engineering the immediate environment relevant to the transition~\cite{purcell1995spontaneous}. The losses induce an irreversible evolution towards the ground state at every time-step (and subsequently pumped towards $|g\rangle - |m\rangle$ subspace by the spin measurement). 

We again use the Kraus representation~\cite{sudarshan_stochastic_1961,jordan_dynamical_1961,kraus_general_1971,kraus1983states} to describe and simulate the unobserved fluorescence dynamics. The Kraus operators describing spontaneous emission between levels $|c\rangle$ and $|g\rangle$ are obtained by extending the framework for describing fluorescence emission of a two-level quantum system discussed in refs.~\cite{lewalle_measuring_2020,jordan_anatomy_2016},

\begin{equation}
    \hat{M}_{c}(0)=\left(
\begin{array}{ccc}
 1 & 0 & 0 \\
 0 & \sqrt{1-\epsilon_{c}} & 0 \\
 0 & 0 & 1 \\
\end{array}
\right),~\hat{M}_{c}(1)=\left(
\begin{array}{ccc}
 0 & 0 & 0 \\
 0 & 0 & 0 \\
 0 & \sqrt{\epsilon_{c}} & 0 \\
\end{array}
\right),
\end{equation}
where $\epsilon_{c}=\gamma_{c}dt$ and $\gamma_{c}$ is the rate of spontaneous emission to the ambient reservoir. Here $0$ and $1$ label no fluorescence emission, and a single fluorescence emission, respectively. The ordering of basis vectors assumed is $(|m\rangle, |c\rangle, |g\rangle)$, such that these Kraus operators represent an amplitude damping channel.
  
  The Kraus operators also satisfy the completeness property, $\sum_{i=0}^{1} \hat{M}_{c}(i)^{\dagger} \hat{M}_{c}(i)=\hat{\mathds{1}}.$ See Appendix~\ref{App1} for a microscopic derivation of the Kraus operators in the spirit of refs.~\cite{jordan_anatomy_2016,lewalle_measuring_2020}. Since losses via fluorescence emission corresponds to unobserved detection events, the contribution to quantum evolution of the three level system from fluorescence emission between levels $c$ and $g$ can be modeled as,
$    d\rho=\hat{M}_{c}(0)\rho\hat{M}^{\dagger}_{c}(0)+\hat{M}_{c}(1)\rho\hat{M}^{\dagger}_{c}(1)
-\rho$. When expanded to first order in $\epsilon_{c}$, this yields the quantum master equation, $d\rho = \epsilon_{c}(L_{c}\rho L_{c}^{\dagger}-\{\rho, L_{c}^{\dagger}L_{c}\}/2)$, where $L_{c}=|g\rangle\langle c|$. Its equivalence to Wigner-Weisskopf description of resonance fluorescence is discussed in~\cite{lewalle_measuring_2020}, considering a quantum two-level system.

\subsection{Observing the clock ticks}
\begin{figure}
    \centering
\includegraphics[width=\linewidth]{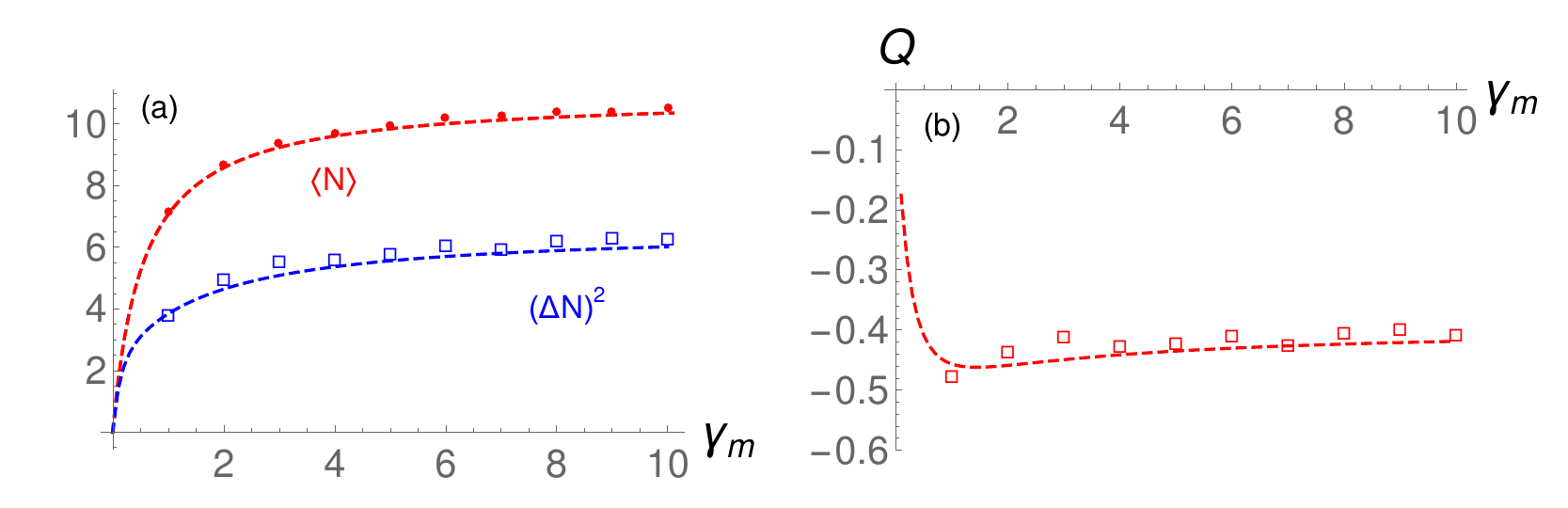}
    \caption{Characterizing the athermal quantum clock in a three level system: (a) Mean and variance of clock ticks. The dashed lines are predictions from using the large deviation principle, discussed in Sec.~\ref{twolevel}. We use $\Omega=1, \omega=0.1 \gamma_{w}=3,\gamma_{c}=4$. In simulations, we generate $5\times 10^3$ number of trajectories using a number of time steps $n=2000$ with $dt=0.5\times10^{-2}$. The system is initialized in the steady state.  (b) The Mandel Q parameter. The dashed lines are predictions based on large deviation principle.}
    \label{fig:athermalthreelevel}
\end{figure}

Clock ticks are produced by transitions between levels $|m\rangle$ and $|c\rangle$.
As discussed in Sec.~\ref{twolevel}, counting the quantum jumps between levels in the clockwork subspace amounts to keeping track of time. 
The quantum jumps in the clockwork subspace are described by the Kraus operators,
\begin{equation}
    \hat{M}_{w}(0)=\left(
\begin{array}{ccc}
  \sqrt{1-\epsilon_{w}} & 0 & 0 \\
 0 &1 & 0 \\
 0 & 0 & 1 \\
\end{array}
\right),~\hat{M}_{w}(1)=\left(
\begin{array}{ccc}
 0 & 0 & 0 \\
 \sqrt{\epsilon_{w}} & 0 & 0 \\
 0 & 0 & 0 \\
\end{array}
\right),
\end{equation}
where $\epsilon_{w}=\gamma_{w}dt$ and $\gamma_{w}$ is the rate of spontaneous emission in the clockwork subspace. 

  The dynamics of the three level system subject to observation of ticks (or no ticks) can be modeled as follows. At any value of coordinate time $t$, a tick is recorded with probability,
\begin{eqnarray}
    p_{\text{tick}}&=&\text{tr}\bigg[\hat{M}_{w}(1)\bigg(\int dr\hat{\mathcal{M}}_{dt}(r)\rho(t)\hat{\mathcal{M}}_{dt}(r)^{\dagger}\bigg)\hat{M}_{w}(1)^{\dagger}\bigg]\nonumber\\&=&\epsilon_{w}\langle m|\bigg(\int dr\hat{\mathcal{M}}_{dt}(r)\rho(t)\hat{\mathcal{M}}_{dt}(r)^{\dagger}\bigg)|m\rangle,
\end{eqnarray}
and no tick is observed with probability, $1-p_{\text{tick}}$. We have denoted $\hat{\mathcal{M}}_{dt}(r)=\hat{M}(r)e^{-i\hat{H}dt}$. Since the observation of ticks is probabilistic ($k=0$ for no ticks, $k=1$ for observing a tick), we have the following conditional evolution of the system at every time $t$,\begin{widetext}
\begin{equation}
    \rho(t+dt) = \frac{\sum_{j=0,1}\hat{M}_{c}(j)\hat{M}_{w}(k)\big[\int dr\hat{\mathcal{M}}_{dt}\rho(t)\hat{\mathcal{M}}_{dt}^{\dagger}\big]\hat{M}_{w}(k)^{\dagger}\hat{M}_{c}(j)^{\dagger}}{\text{tr}\big\{\hat{M}_{w}(k)\big[\int dr\hat{\mathcal{M}}_{dt}\rho(t)\hat{\mathcal{M}}_{dt}^{\dagger}\big]\hat{M}_{w}(k)^{\dagger}\big\}}.
\end{equation}
   \end{widetext}
    In simulations, we approximate $\int dr\hat{\mathcal{M}}_{dt}\rho(t)\hat{\mathcal{M}}_{dt}^{\dagger}\approx\rho(t)-dt \big(i[\hat{H},\rho]+\gamma_{m}[\hat{X},[\hat{X},\rho(t)]]\big)$.

We now proceed to characterize the statistics of total number of ticks observed within a given large but finite time period 
using the the principle of large deviation.
 
\subsection{The statistics of ticks}
To derive the statistics of ticks, we again follow the large deviation principle. The tilted Lindblad generator is given by,
\begin{eqnarray}
    \mathcal{W}(s)[\rho] &=& -i[\hat{H},\rho]-\gamma_{m}[\hat{X},[\hat{X},\rho]]\nonumber\\&+&\gamma_{c}(L_{c}\rho L_{c}^{\dagger}-\{\rho, L_{c}^{\dagger}L_{c}\}/2)\nonumber\\&+&e^{-s}\gamma_{w}L_{w}\rho L_{w}^{\dagger}-\gamma_{w}\{\rho, L_{w}^{\dagger}L_{w}\}/2,
\end{eqnarray}
where $L_{w}=|c\rangle\langle m|$ and $L_{c}=|g\rangle\langle c|$.

The moment generating function for the statistics of ticks is the largest real eigenvalue of $\mathcal{W}$. Please refer to supplemental Mathematica~\cite{Mathematica} files for details of how we extract this~\footnote{The associated Mathematica notebook can be found in the following \href{https://github.com/sreenathkm92/clocks.git}{ Github repository link}.}.  The steady state can be computed from solving $\frac{d\rho}{dt}=\mathcal{W}(0)[\rho] = 0$.   Following an analysis similar to Sec.~\ref{twolevel}, we find that the clock ticks at a rate
$\gamma_{\text{tick}}=P_{m}(\theta)\gamma_{w}$, where $P_{m}$ is the steady state occupation of level $|m\rangle$ (see App.~\ref{App2M}). For $\theta = \pi/2$, this is given by
\begin{equation}
\gamma_{\text{tick}}=\frac{2 \gamma_{c} \gamma_{m}\gamma_{w}}{\gamma_{c} (4 \gamma_{m}+\gamma_{w})+2 \gamma_{m} \gamma_{w}},
\end{equation} 
and the average number of ticks in a given duration $t$ is given by $\langle N\rangle = \gamma_{\text{tick}}t$. For large but finite $t$, this agrees with the predictions based on the large deviation principle [see Fig.~\ref{fig:athermalthreelevel}(a)]. We also compute the variance of ticks based on simulations as well as the large deviation principle, where we find that the statistics is generically sub-Poissonian with the Mandel parameter $Q<0$. Their agreement is shown in Fig.~\ref{fig:athermalthreelevel}(b). The energy converted into clockwork in unit time is $J_{w}=\gamma_{\text{tick}}\Omega_{w}$ in the steady state.

Conversely, given the number of counts $N$, one can estimate the working duration $t_{w}$ the clock has passed as $t_{w}=N/\gamma_{\text{tick}}$. The accuracy of the clock in terms of the relative error in the number of counts $N$ for a given working duration $t_{w}$ is given by (for $\theta=\pi/2$),
\begin{equation}
    \frac{\delta N(t_{w})}{\langle N(t_{w})\rangle}=\frac{\sqrt{1+Q}}{\sqrt{\gamma_{\text{tick}}t_{w}}}=\sqrt{\frac{(1+Q)[\gamma_{c} (4 \gamma_{m}+\gamma_{w})+2 \gamma_{m} \gamma_{w}]}{2 \gamma_{c} \gamma_{m}\gamma_{w}t_{w}}}.
\end{equation}
The clock functions better with increasing the duration $t_{w}$ we intend to measure, and has a better accuracy than typical radio-carbon clocks~\cite{milburn_thermodynamics_2020} which obey a Poissonian statistics for the total number of counts within a given time period.

    It is also of pedagogical interest to estimate the number of counts $N(t_w)=\gamma_{\text{tick}}t_{w}$ when the relative error $\delta N(t_w)$ is of the order of a single tick, i.e., when $\delta N(t_w)\geq 1$. This would be another measure for the accuracy of the clock as in~\cite{erker_autonomous_2017}. It is straightforward to see that $\delta N(t_w)\geq 1$ when $N_w\geq (1+Q)^{-1}$. This suggests that for Poissonian clocks ($Q=0$), the uncertainty of counts becomes of order one after a single tick itself. It is interesting to note that a negative Mandel's $Q$ factor, $-1\leq Q<0$ therefore offers a sub-Poissionian improvement to the accuracy of a clock, which however is only marginal for the examples presented here. We may also compare this to the accuracy of a thermal clock as defined in~\cite{erker_autonomous_2017}; For a $d$ dimensional register, the accuracy of a thermal clock is given by $d\times$ the population bias, $d\tanh{(\beta\Omega_{w}/2)}<d$, where $\beta^{-1}=k_{\text{B}}T_{\text{eff}}$ where $T_{\text{eff}}$ is the effective temperature driving the clock transition. Since the clock register we consider involves only one transition in our examples we may assume $d=1$ for comparison, in which case we see that $Q<0$ has an advantage. However, it remains to be tested if increasing the register dimension improves the clock's accuracy for the examples discussed in the present manuscript.

\subsection{Quantumness of optimal clocks}
In addition to the observed statistics of ticks, we wish to highlight that the fueling mechanism for the athermal clock discussed here is of quantum mechanical nature, as some degree of non-commutativity between the clock's Hamiltonian and the measured observable is necessary to make the clock tick. For generic measurement angles $\theta$, the steady state has quantum coherences between levels $|g\rangle$ and $|m\rangle$, which is periodic in $2\theta$, and vanishing for $\theta=0, \pi/2, \pi$ (see Appendix.~\ref{App2}). Therefore our clock model also presents interesting trade-offs between the degree of coherence in the steady state $|\rho_{mg}|^{2}$, and the ticking rate of the clock. A point worth emphasizing is that the fueling mechanism for the optimal clock realization is also strictly quantum even though the clock does not create any quantum coherences in the steady state when $\theta=\pi/2$. This is because the measurement interactions when $\theta=\pi/2$ couples the apparatus to the transition dipole between levels $|g\rangle$ and $|m\rangle$, and similar to generic measurement angle $\theta$, this is a unitary rotated basis compared to the clock's Hamiltonian. Such considerations are not feasible in a classical system~\cite{jacobs_how_2003}, making the resources fueling the clock strictly quantum.

\section{Hybrid quantum clocks\label{hybrid}}
We now look at possible generalizations of our model to including thermal resources in parallel to athermal resources for driving ticks of the quantum clock. This for example, allows to compare our model to the thermal quantum clock discussed in~\cite{erker_autonomous_2017}, and explore the effect of observing ticks in the dynamics of the clock itself.
The objective is to describe scenarios where the transition $|g\rangle\rightarrow |m\rangle$ is fueled by both thermal and measurement processes by including a hot reservoir that selectively couples to the $|g\rangle\rightarrow |m\rangle$ transition. We also replace the unobserved fluorescence emission from $|c\rangle\rightarrow |g\rangle$ (equivalent to coupling to a zero temperature bath) with coupling to a reservoir at a cold but finite temperature. We consider weakly coupled reservoirs, for which the statistics of ticks can be computed using a tilted Lindbladian of the following form:
\begin{eqnarray}
    \mathcal{W}(s)[\rho] &=& -i[\hat{H},\rho]-\gamma_{m}[\hat{X},[\hat{X},\rho]]+\sum_{k=h,c}\mathcal{D}_{k}(\rho)\nonumber\\&+&\gamma_{w}(e^{-s}L_{w}\rho L_{w}^{\dagger}-\{\rho, L_{w}^{\dagger}L_{w}\}/2),
\end{eqnarray}
where $\mathcal{D}_{k}(\rho)=\gamma_{k}(L_{k}\rho L_{k}^{\dagger}-\{\rho, L_{k}^{\dagger}L_{k}\}/2)+\gamma_{k}e^{-\beta_{k}\Omega_{k}}(L_{k}^{\dagger}\rho L_{k}-\{\rho, L_{k}L_{k}^{\dagger}\}/2)$~\cite{breuer2002theory}. We denote $L_{h}=|g\rangle\langle m|~,L_{c}=|g\rangle\langle c|,$ $\Omega_{h}=\Omega_{m}$, and $L_{w}=|c\rangle\langle m|$.

\begin{figure}
    \centering
\includegraphics[width=\linewidth]{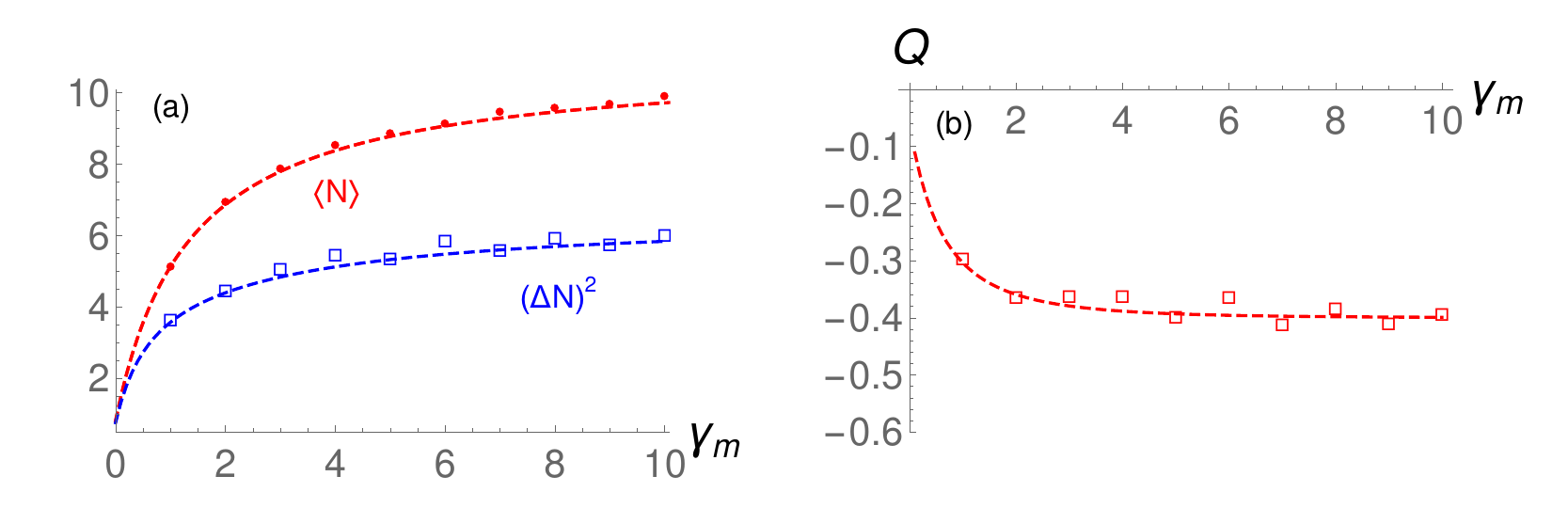}
    \caption{Characterizing the hybrid quantum clock in a three level system: (a) Mean and variance of clock ticks. The data points are based on simulations. The dashed lines are predictions from using the large deviation principle. We use $\Omega_{m}=1, \Omega_{c}=0.1, \theta=\pi/2, \phi = 0, \gamma_{h}=4, \gamma_{w}=3$, $\beta_{h}\Omega_{m}=3$, $\beta_{c}\Omega_{c}=100$, and $\gamma_{c} = 4$. In simulations, we generate $5\times 10^3$ number of trajectories using a number of time steps $n=2000$ with $dt=0.5\times 10^{-2}$. The system is initialized in the ground state.  (b) The Mandel Q parameter. The dashed lines are predictions based on large deviation principle.}
    \label{fig:hybrid1}
\end{figure}

The steady state (which is solution to $\frac{d\rho}{dt}=\mathcal{W}(0)[\rho] = 0$) of the clock naturally generalizes as a result of including all reservoirs. The mean and variance of ticks can be estimated from the moment generating function for total number of ticks within a given time $t$, which is estimated as the largest real eigenvalue of $\mathcal{W}$. Please refer to the supplemental Mathematica~\cite{Mathematica} notebook for details of how we compute this.  The average number of ticks for finite but large times $t$ is given by $\langle N\rangle = P_{m}\gamma_{w}t$, where $P_{m}$ population of the excited state in the steady state. The variance is estimated similarly, which shows sub-Poissonian statistics, indicating the quantum nature of the clock. See Fig.~\ref{fig:hybrid1}, where we compare the predictions of large deviation theory for the measured statistics with numerical simulations.

Before we conclude, we wish to briefly comment on possible experimental realizations of the clocks discussed here. One may consider two different scenarios for the experimental implementation of our proposal: (1) a condensed matter realization where the measurements are performed by a steady stream of electrons through a point contact, and (2) an optical realization, where measurements are performed with a steady stream of photons through a wave-guide. Both invite considerable interest in exploring further, the fundamental limitations on generating clock ticks using athermal resources. The fluorescence emission rate can be related to the inverse of the coherence time of quantum two level, and three level systems, which makes the clock tick at microsecond intervals or faster in a superconducting circuit platform. Superconducting circuit serves as the immediate platform where the proposal can be investigated experimentally~\cite{weber_mapping_2014,vijay_stabilizing_2012,minev_catch_2019}, also given the excellent measurement and control capabilities achieved in a three level quantum system architecture~\cite{minev_catch_2019}. Comparable quantum clock models have also been demonstrated recently in a superconducting platoform~\cite{he_measurement_2022}. In addition, simultaneous time-continuous measurements of both spin and fluorescence have also been realized~\cite{ficheux_dynamics_2018}, which closely meets the requirements for experimentally implementing our proposal. Finally, Purcell enhancement~\cite{purcell1995spontaneous} of the clock transition maybe feasible which may also improve the performance of the clock.

\section{Discussions\label{discuss}}We presented simple models for engineering non-equilibrium conditions in a three level quantum system by combining different quantum measurements---of spin and fluorescence---treated as athermal resources. 
A quantum clock realization was the focus of this article where the statistics of the total number of ticks within a given time period are generically sub-Poissonian, alluding to the quantum nature of observed ticks. We used the large deviation principle for resonance flourescence and the Mandel's Q parameter to quantify the statistics.   Our approach differ from existing literature on applying large deviation principle to study the flourescence counting statistics~\cite{garrahan_thermodynamics_2010}, owing to the inclusion of athermal resources to drive the fluorescence ticks.   Considering simple examples, we also obtain closed form results for the moment generating function that determines the full counting statistics in the limit of large but finite time.

A natural consequence of the sub-Poissonian statistics obtained is that the clock ticks more regularly than a radio-carbon clock, which obeys a Poissonian statistics~\cite{milburn_thermodynamics_2020}.   We also briefly discussed the sense in which the fueling of athermal clocks is strictly quantum. Certain degree of non-commutativity between the measured observable and the clock's Hamiltonian is necessary to fuel the clock ticks, and an optimal clock is realized in our description when the measured observable is maximally non-commuting with the clock's Hamiltonian. Such considerations are strictly quantum, as one requires to measure in a unitary rotated basis (as opposed to the clock's Hamiltonian) to fuel the clock, which is not feasible in a classical three level system. The clock ticks even at zero temperature, consuming resources such as chemical potential gradients driving the measurement apparatuses, as opposed to temperature gradients in a thermal quantum clock. The non-equilibrium resources realize a quantum battery~\cite{campaioli_enhancing_2017,barra_dissipative_2019,binder_quantacell_2015,levy_quantum_2016}  in effect, fueling the ticks of the quantum clock. 

We also extended our analysis to include hybrid quantum clocks, where both thermal and athermal resources are used to fuel the clock. We find that observing the clock ticks has non-trivial back-action on the ticking rate of the clock for such hybrid clocks as well; the non-equilibrium steady state populations which determines the rate at which the clock ticks is strictly athermal, even when only thermal resources are consumed. 

We would also like to bring the reader's attention to a relatable recent work~\cite{he_measurement_2022}, where the authors propose and experimentally demonstrate a quantum two level system based quantum clocks driven by measurements. In their work, the two level quantum system is driven along the $x$ axis by a laser field, and the clock signal is monitored via the homodyne quantum measurements of spin along $y$ axis to realize an oscillatory quantum clock. A non-oscillatory clock is also discussed, where the qubit's energy is measured in the strong measurement limit. We note that while the motivation for the present work is indeed similar, we addressed a different class of measurement fueled, and hybrid quantum clocks. The main difference is that the clock model discussed here in a three level quantum system treats the resource fueling the clocks [unobserved spin (in $|m\rangle$---$|g\rangle$ subspace), and fluorescence (in $|c\rangle$---$|g\rangle$ subspace)] different from the observed clock signal (fluorescence measurements between levels $|m\rangle$ and $|g\rangle$). In comparison, the clock signal in~\cite{he_measurement_2022} results from the measurement records of spin which also drives the clock, together with the laser field. This difference is also related to the key highlight of the present work, that ignorant observers can (be considered as resource to) make a quantum clock tick. The statistics of clock ticks we obtained is generically sub-Poissonian, which also differs from the non-oscillatory Poissonian clocks in Refs.~\cite{he_measurement_2022,milburn_thermodynamics_2020}.

It is also worthwhile to re-iterate that the mathematical key distinction to thermal clocks (for example, discussed in~\cite{erker_autonomous_2017}) originates from the quantum measurement fueling process, introducing Hermitian dissipators in the problem. 
When some of the Lindblad operators are Hermitian, i.e., $L_{k}^{\dagger}=L_{k}=\sqrt{2}X_{k}$, we can split the Lindblad master equation to non-Hermitian (c) and Hermitian (r) contributions,
\begin{eqnarray}
    \frac{\partial\rho}{\partial t} &=& -i[\hat{H}(t),\rho]+\sum_{c}\gamma_{c}(L_{c}\rho L_{c}^{\dagger}-\{\rho, L_{c}^{\dagger}L_{c}\}/2)\nonumber\\&-&\sum_{r}\gamma_{r} [X_{r},[X_{r},\rho]].
\end{eqnarray}
The thermal contributions, usually discussed for thermal clocks and other thermal quantum machines comes from the non-hermitian Lindblad operators $L_{k}$, while the clocks here leverage on the additional terms feasible above that contain Hermitian $L_{k}$s to fuel the clock ticks. In that spirit, we referred to the Hermitian $L_{k}$ contributions as athermal. As we discussed, such contributions can be related to underlying athermal resources such as chemical potential gradients in the system, which sets the necessary non-equilibrium conditions.

\begin{figure}[t]
    \includegraphics[width=0.9\linewidth]{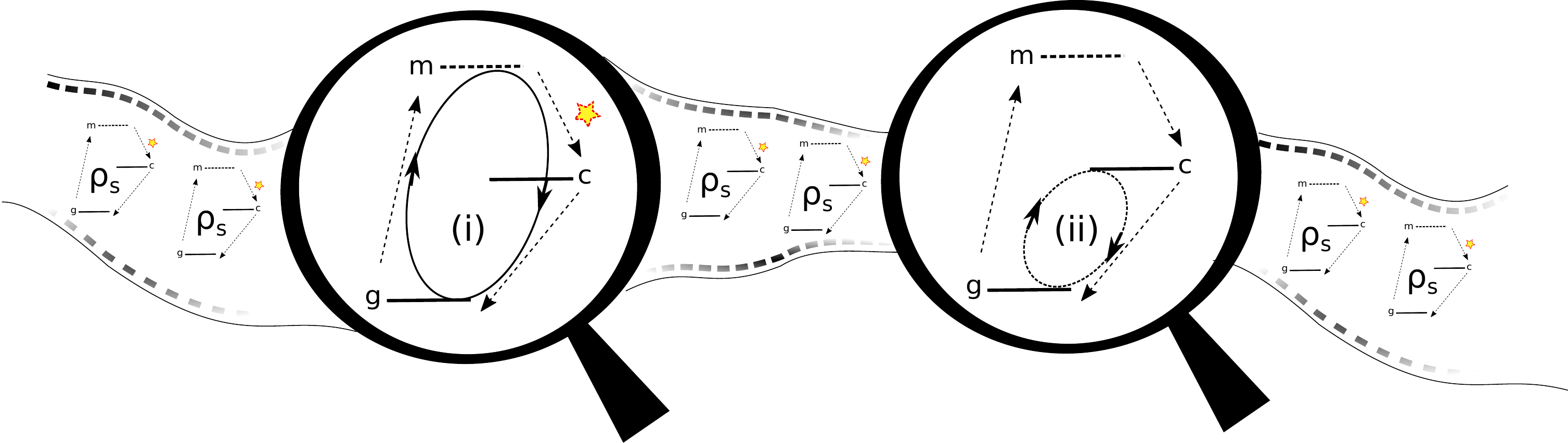}
    \caption{Different cycles possible in a three level quantum system acting as a clock device: (i) A clock cycle that contributes to generate clock ticks. (ii) A dissipative cycle, where the energy contributed by measurements is dissipated as heat.} 
    \label{fig:diss}
\end{figure}

The methods we used for realizing non-equilibrium steady states in a simple quantum system by combining different types quantum measurements can have exciting device applications beyond the quantum clock realization we focused in the present article. The model can be generalized in a straightforward manner to include different observers who measure observables of the clock which do not commute. It may also be generalized to incorporate different observers watching the quantum clock tick, towards realizing observer dependent clocks in the quantum sense.  The two and three level quantum systems acting as transducers for generating the ticks of the quantum clock may also be reverse engineered to use as a probe for characterizing the underlying temperature, and chemical potential gradients, leading to interesting quantum sensor applications for the device.

On the fundamental side, the present work mostly focused on the large time behavior of the clock. An interesting regime to look further into, particularly in view of alternate quantum device implementations, would be the transient regime. Our model with a three level quantum system is simple enough that the finite-time statistics of relevant thermodynamic variables can be derived exactly in different regimes of operation, accounting for individual realizations of the quantum measurement process.  The measurements which fuel the ticks enable different types of cycles in the three level quantum system in its steady state, $\rho_{s}$. This can be noticed by, for example, studying the strong measurement limit $\gamma_{m}\gg\gamma_{c},\gamma_{w}$. In this limit, the following cycles maybe observable upon resolving the individual realizations of quantum spin and fluorescence measurement processes (see Fig.~\ref{fig:diss}): 
\begin{itemize}
    \item[(i)] A cycle that generates clock ticks. Occurs when the spin measurement outcomes are polarized towards the subspace $\mathds{1}-|c\rangle\langle c|$. The cycle evolves as,  $\rho_{s} \rightarrow ...\rightarrow (\text{states polarized towards } |\theta,\phi,+\rangle\langle\theta,\phi,+|$ or $|\theta,\phi,-\rangle\langle\theta,\phi,-|$) $\rightarrow |c\rangle \rightarrow |g\rangle\rightarrow ...\rightarrow\rho_{s}$. 
    \item[(ii)] A fundamentally dissipative cycle. Occurs when the spin measurement outcomes are polarized towards the $|c\rangle$ state. The cycle evolves as, $\rho_{s} \rightarrow ... \rightarrow$ states polarized towards $|c\rangle \rightarrow |g\rangle\rightarrow ...\rightarrow\rho_{s}$. \end{itemize}
 The presence of the dissipative cycle ensures that a part of the energy given by the measurement is necessarily dissipated into the ambient zero temperature reservoir as heat, without contributing to the clockwork. The statistical likelihood of occurrence of these cycles therefore offers insights into the fundamental limitations of observing clock cycles and time-periodic behaviours in a three level quantum system. Keeping track of these cycles also points at the feasibility of oscillatory quantum clocks using engineered reservoirs, comparable to the oscillatory clocks proposed in~\cite{he_measurement_2022}. Characterizing thermodynamic cycles in steady state thermoelectric devices have also become a subject of significant interest in recent years~\cite{PhysRevB.103.075404}.  Additionally, the role of quantum coherence in generic, finite-time clock cycles, and the effect of making quantum mechanical observations on the cycles themselves are also interesting directions worth exploring further. We defer such analyses to a future work.
\section{Acknowledgments} The author thanks Cyril Elouard for helpful suggestions on the manuscript. The author also thanks anonymous referee(s) for helpful comments on the statistics presented in the manuscript.  This work was supported by the Wallenberg Initiative on Networks and Quantum Information (WINQ). Nordita is partially supported by Nordforsk.

\noindent\textit{Data availability statement---}The simulations associated to figures in the manuscript can be found in the following \href{https://github.com/sreenathkm92/clocks.git}{ Github repository link}. 
\appendix
\begin{widetext}
 
\section{Kraus operators for spin measurements\label{KrausSpin}}
Here we provide a pedagogical derivation of Kraus operators for spin measurements in two and three level systems considered in the manuscript, based on the standard measurement model where the system interacts with a measurement apparatus unitarily, and the measurement apparatus is subsequently projected onto the readout basis. For this, we consider a continuous variable measurement apparatus (photons, or quantum LC circuits) in a Gaussian initial state~\cite{jackson_how_2021,jacobs_straightforward_2006},
\begin{equation}
    |\psi_{M}\rangle =[\pi/(4\gamma_{m})]^{-\frac{1}{4}} \int dx e^{-2\gamma_{m} x^{2}}|x\rangle.
\end{equation}
The interaction Hamiltonian couples the spin observable $\hat{X}$ to the momentum of the measurement apparatus via the following interaction Hamiltonian,
\begin{equation}
H_{\text{int}}dt=\sqrt{dt}\hat{p}\hat{X}.
\end{equation}
We imagine that the measurement is a homodyne detection of the quadrature $\hat{x}$ (conjugate to $\hat{p}$) of the continuous variable probe. This could be the position quadrature of a photon, or the charge of a quantum LC circuit. The measurement operator is then obtained by partial projection for a given readout $x=q$,
\begin{equation}
    \hat{M}_{\hat{X}}(q)=\langle q|e^{-iH_{\text{int}}dt}|\psi_{M}\rangle = [\pi/(4\gamma_{m})]^{-\frac{1}{4}} \exp{\bigg[-2\gamma_{m}(q\mathbb{I}-\hat{X}\sqrt{dt})^{2}\bigg]}.
\end{equation}
The Homodyne readout is scaled as $r=q/\sqrt{dt}$, and in terms of $r$, we can write the measurement operator as,
\begin{equation}
    \hat{M}_{\hat{X}}(r)= [\pi/(4\gamma_{m}dt)]^{-\frac{1}{4}} \exp{\bigg[-2\gamma_{m}dt(r\mathbb{I}-\hat{X})^{2}\bigg]}=(4\epsilon_{m}/\pi)^{\frac{1}{4}} \exp{\bigg[-2\epsilon_{m}(r\mathbb{I}-\hat{X})^{2}\bigg]},
\end{equation}
where we denote $\epsilon_{m}=\gamma_{m}dt$.

\section{Kraus operators for resonance fluorescence emission from a three level system\label{App1}}
Here we closely follow the derivation of Kraus operators for resonant fluorescence via photon counting of a two level system presented in~\cite{lewalle_measuring_2020} and extend it to a three level quantum system. We consider the an arbitrary initial state for the three level system, $|\psi_{0}\rangle=\alpha_{g} |g\rangle +\alpha_{m}|m\rangle+\alpha_{c}|c\rangle,$ and the electromagnetic field mode that couple to the atomic levels c--g in its ground state, $|0_{c}\rangle.$ To small time $dt$, the combined evolution of the three level system and the field mode can be written as,
\begin{eqnarray}
    |\psi_{dt}\rangle &=& \big(\alpha_{g}|g\rangle+\alpha_{m}|m\rangle)|0_{c}\rangle+\alpha_{c}\big(\sqrt{1-\epsilon_{c}})|c0_{c}\rangle+\sqrt{\epsilon_{c}}|g1_{c}\rangle\big)\nonumber\\&=& (\alpha_{g}|g\rangle+\alpha_{m}|m\rangle)+\alpha_{c}\big(\sqrt{1-\epsilon_{c}})|c\rangle+\sqrt{\epsilon_{c}}a_{c}^{\dagger}|g\rangle\big)|0_{c}\rangle\nonumber\\
    &=&\big(|g\rangle\langle g|\otimes\mathds{1}_{c}+|m\rangle\langle m|\otimes\mathds{1}_{c}+\sqrt{1-\epsilon_{c}})|c\rangle\langle c|\otimes\mathds{1}_{c}+\sqrt{\epsilon_{c}}a_{c}^{\dagger}|g\rangle\langle c|\big)|\psi_{0}\rangle|0_{c}\rangle\nonumber.
\end{eqnarray}
The Kraus operator evolution corresponding to no-detection event is,
\begin{equation}
\hat{M}_{c}(0)|\psi_{0}\rangle=\langle 0_{c}     |\psi_{dt}\rangle = \big(|g\rangle\langle g|+|m\rangle\langle m|+\sqrt{1-\epsilon_{c}}|c\rangle\langle c|\big)|\psi_{0}\rangle,
\end{equation}
and the Kraus operator evolution corresponding to single photon detection event is,
\begin{equation}
\hat{M}_{c}(1)|\psi_{0}\rangle=\langle 1_{c}     |\psi_{dt}\rangle =\big( \sqrt{\epsilon_{c}}|g\rangle\langle c|\big)|\psi_{0}\rangle.
\end{equation}
It is straight-forward to verify that $\hat{M}_{c}(0)^{\dagger} \hat{M}_{c}(0)+\hat{M}_{c}(1)^{\dagger} \hat{M}_{c}(1)=\hat{\mathds{1}}_{3\times 3}.$ Fluorescence emission corresponds to unobserved detection events, so the quantum evolution of the three level system subject to fluorescence emission between levels $c$ and $g$ can be modeled as,
$    \rho(t+dt)=\hat{M}_{c}(0)\rho(t)\hat{M}^{\dagger}_{c}(0)+\hat{M}_{c}(1)\rho(t)\hat{M}^{\dagger}_{c}(1).
$
\section{The steady state of the three level quantum clock\label{App2M}}

In order to compute the steady state, we may use the unconditional evolution of the density matrix of the three level quantum system dictated by,
\begin{equation}
    \frac{\partial\rho}{\partial t} = -i[\hat{H},\rho]-\gamma_{m}[\hat{X},[\hat{X},\rho]]+\mathcal{M}_{c}(\rho)+\mathcal{M}_{w}(\rho),
\end{equation}
where $\mathcal{M}_{w}(\rho)=\gamma_{w}(L_{w}\rho L_{w}^{\dagger}-\{\rho, L_{w}^{\dagger}L_{w}\}/2)$, $\mathcal{M}_{c}(\rho)=\gamma_{c}(L_{c}\rho L_{c}^{\dagger}-\{\rho, L_{c}^{\dagger}L_{c}\}/2)$, $L_{c}=|g\rangle\langle c|$, and $L_{w}=|c\rangle\langle m|$ describes the dynamics of the clock as a result of averaging over the clock ticks. The density matrix elements evolve as,
\begin{eqnarray}
\frac{d\rho_{mm}}{dt}&=& 
 2 \gamma_{m}[\text{Re}(\rho_{mg}) \cos(\phi) + \text{Im}(\rho_{mg}) \sin(\phi)] \sin(2 \theta)\nonumber\\&+& ( \rho_{mm}-\rho_{gg}) \gamma_{m}[\cos(2 \theta)-1]- 
 \rho_{mm} \gamma_{w} , \nonumber\\
 \frac{d\rho_{cc}}{dt}&=&\rho_{mm} \gamma_{w}-\rho_{cc} \gamma_{c},\nonumber\\
 \frac{d\rho_{mg}}{dt} &=&- \rho_{mg} [2 \gamma_{m} \cos (2 \theta)+6 \gamma_{m}+\gamma_{w}+2 i \Omega_{c} ]/2\nonumber\\&+& \gamma_{m} e^{i \phi} ( \rho_{mm}-\rho_{gg}) \sin (2 \theta)+2 \gamma_{m} e^{2 i \phi} \rho_{gm} \sin ^2(\theta),\nonumber\\
 \frac{d\rho_{mc}}{dt}&=&-(\rho_{mc}/2) (\gamma_{c}+2 \gamma_{m}+\gamma_{w}-2 i \Omega_{c} +2 i \Omega_{m} ),\nonumber\\
 \frac{d\rho_{cg}}{dt}&=&-(\rho_{cg}/2) (\gamma_{c}+2 \gamma_{m}+2 i \Omega_{c}).
 \label{eqns}
\end{eqnarray}
The above dynamical equations are written in terms of a coordinate time $t$, while the clock properties we investigate are strictly restricted to the steady state $d\rho/dt = 0$ to have minimal or no dependence on the coordinate time.  The population of levels $m$ and $c$ in the steady state are given by (when $\theta=\pi/2,~\forall\phi$),
\begin{eqnarray}
P_{m}&=&\frac{2 \gamma_{c} \gamma_{m}}{\gamma_{c} (4 \gamma_{m}+\gamma_{w})+2 \gamma_{m} \gamma_{w}}, \nonumber\\  
    P_{c}&=&\frac{2 \gamma_{w} \gamma_{m}}{\gamma_{c} (4 \gamma_{m}+\gamma_{w})+2 \gamma_{m} \gamma_{w}}.\nonumber\\
\end{eqnarray}
More general expressions for arbitrary $\theta$ are given in Appendix~\ref{App2}, where we note that the population $P_{m}$ is maximized when $\theta=\pi/2$, i.e., when the measured spin observable maximally non-commutes with the clock's Hamiltonian. For finite $\gamma_{m}$, we can also define an effective temperature $T_{w}$ for the clockwork subspace in the steady state, given by,
\begin{equation}
e^{-\beta_{w} \Omega_{w}}=\frac{P_{m}}{P_{c}}=\frac{\gamma_{c}}{\gamma_{w}},    
\end{equation}
where $\beta_{w}=(k_{\text{B}}T_{w})^{-1}$. This suggests a fundamental limitation of the clock that population inversion (or negative effective temperature $T_{w}$) requires a larger $\gamma_{c}$ compared to $\gamma_{w}$. On the other hand, quantum observations of ticks makes the clock tick even in the absence of a population inversion; watching a quantum clock tick, makes it tick!

\subsection{arbitrary $\theta$\label{App2}}
The population of levels $m$ and $c$ in the steady state for arbitrary measurement angle $\theta$ ($\forall\phi$) are given by (in the absence of thermal resources),
\begingroup\makeatletter\def\f@size{8}\check@mathfonts
\begin{eqnarray}
P_{m}&=&\frac{2 \gamma_{c}\gamma_{m}\sin ^2(\theta) \left(8 \gamma_{m}\gamma_{w}+\gamma_{w}^2+4 \Omega_{m}^2\right)}{\gamma_{m}\cos (2 \theta) \left(\gamma_{w}(2 \gamma_{c}-\gamma_{w}) (8 \gamma_{m}+\gamma_{w})-4 \Omega_{m}^2 (2 \gamma_{c}+\gamma_{w})\right)+4 \Omega_{m}^2 (\gamma_{w}(\gamma_{c}+\gamma_{m})+2 \gamma_{c}\gamma_{m})+\gamma_{w}(8 \gamma_{m}+\gamma_{w}) (\gamma_{w}(\gamma_{c}+\gamma_{m})+6 \gamma_{c}\gamma_{m})},\nonumber\\  
    P_{c}&=&-\frac{2 \gamma_{m}\gamma_{w}\sin ^2(\theta) \left(8 \gamma_{m}\gamma_{w}+\gamma_{w}^2+4 \Omega_{m} ^2\right)}{\gamma_{m}\cos (2 \theta) \left(\gamma_{w}(\gamma_{w}-2 \gamma_{c}) (8 \gamma_{m}+\gamma_{w})+4 \Omega_{m}^2 (2 \gamma_{c}+\gamma_{w})\right)-4 \Omega_{m}^2 (\gamma_{w}(\gamma_{c}+\gamma_{m})+2 \gamma_{c}\gamma_{m})-\gamma_{w}(8 \gamma_{m}+\gamma_{w}) (\gamma_{w}(\gamma_{c}+\gamma_{m})+6 \gamma_{c}\gamma_{m})}.\nonumber\\
\end{eqnarray}
 \endgroup

\noindent The degree of coherence between levels $|g\rangle$ and $|m\rangle$ is given by,
\begingroup\makeatletter\def\f@size{7}\check@mathfonts
\begin{equation}
|\rho_{mg}|^{2}=\frac{4 \gamma_{c}^2 \gamma_{m}^2 \gamma_{w}^2 \sin ^2(2 \theta) \left((8 \gamma_{m}+\gamma_{w})^2+4 \Omega_{m} ^2\right)}{\left(\gamma_{m} \cos (2 \theta) \left(\gamma_{w} (2 \gamma_{c}-\gamma_{w}) (8 \gamma_{m}+\gamma_{w})-4 \Omega_{m}^2 (2 \gamma_{c}+\gamma_{w})\right)+4 \Omega_{m}^2 (\gamma_{w} (\gamma_{c}+\gamma_{m})+2 \gamma_{c} \gamma_{m})+\gamma_{w} (8 \gamma_{m}+\gamma_{w}) (\gamma_{w} (\gamma_{c}+\gamma_{m})+6 \gamma_{c} \gamma_{m})\right)^2}.\\
\end{equation}
 \endgroup
The other density matrix elements are equal to zero in the steady state. The above results generalize to when thermal resources are also consumed, in a straightforward manner. 

\subsection{Steady state of a hybrid quantum clock}
Following a similar analysis for hybrid quantum clocks in a three level system, we find that the average population of the excited state $|m\rangle$ in the steady state is given by,
\begin{equation}
  P_{m}(\theta=\pi/2)=\frac{\gamma_{c} e^{\beta_{c} \Omega_{c} } \left(2 \gamma_{m} e^{\beta_{h} \Omega_{m} }+\gamma_{h}\right)}{e^{\beta_{c} \Omega_{c} } \left(e^{\beta_{h} \Omega_{m} } (2 \gamma_{m} (2 \gamma_{c}+\gamma_{w})+\gamma_{c} (\gamma_{h}+\gamma_{w}))+\gamma_{h} (\gamma_{c}+\gamma_{w})\right)+\gamma_{c} e^{\beta_{h} \Omega_{m} } (2 \gamma_{m}+\gamma_{h}+\gamma_{w})},  
\end{equation}
and
\begin{equation}
   P_{m}(\theta=\pi)= \frac{\gamma_{c} \gamma_{h} e^{\beta_{c} \Omega_{c} }}{\gamma_{c} (\gamma_{h}+\gamma_{w}) e^{\beta_{c} \Omega_{c} +\beta_{h} \Omega_{m} }+\gamma_{h} e^{\beta_{c} \Omega_{c} } (\gamma_{c}+\gamma_{w})+\gamma_{c} e^{\beta_{h} \Omega_{m} } (\gamma_{h}+\gamma_{w})}.
\end{equation}
A crucial observation is that the steady state remains athermal even in the absence of spin measurements. This can be seen, for example by looking at the effective inverse temperature in the clockwork subspace (when $\theta=\pi$):
\begin{equation}
    e^{-\beta_{w} \Omega_{w}}=\frac{P_{m}}{P_{c}}=\frac{\gamma_{c} \gamma_{h} e^{\beta_{c} \Omega_{c} }}{\gamma_{h} \gamma_{w} e^{\beta_{c} \Omega_{c} }+\gamma_{c} e^{\beta_{h} \Omega_{m} } (\gamma_{h}+\gamma_{w})}.
\end{equation} 
The populations generically differ from the non-equilibrium steady state of a three level quantum system coupled to different temperatures. This is because observing the clock ticks modifies the asymptotic behavior of the clock, which is an additional insight we gain from modeling the tick observations quantum mechanically. In comparison to our earlier examples clocks fueled by measurements only, the hybrid clock can produce ticks even when $\theta=0,~\pi$, assisted by thermal resources. 
   
 \end{widetext}

\bibliography{clocks.bib}
		   \end{document}